\documentstyle[11pt,newpasp,twoside]{article}

\def\edcomment#1{\iffalse\marginpar{\raggedright\sl#1\/}\else\relax\fi}
\reversemarginpar

\begin{document}
\title{Hipparcos Luminosities and Asteroseismology}
\author{Timothy R. Bedding}
\affil{School of Physics, University of Sydney 2006, Australia}

\begin{abstract}
Asteroseismology involves using the resonant frequencies of a star to infer
details about its internal structure and evolutionary state.  Large efforts
have been made and continue to be made to measure oscillation frequencies
with both ground- and space-based telescopes, with typical precisions of
one part in $10^3$--$10^4$.  However, oscillation frequencies are most
useful when accompanied by accurate measurements of the more traditional
stellar parameters such as luminosity and effective temperature.  The
Hipparcos catalogue provides luminosities with precisions of a few percent
or better for many oscillating stars.  I briefly discuss the importance of
Hipparcos measurements for interpreting asteroseismic data on three types
of oscillating stars: $\delta$~Scuti variables, rapidly oscillating Ap
stars and solar-like stars.
\end{abstract}

\section{Preamble: To `e' or not to `e'?}

Because my talk was the last of the day, I chose to begin it with some
controversy.  I suggested that we heed the urgings of Trimble (1995) to
adopt the spelling `astroseismology.'  My basis was a search using the ADS
Abstract Service\footnote{\tt
http://adsabs.harvard.edu/abstract\_service.html} that showed this spelling
in the titles of the first three papers on the subject, by
Christensen-Dalsgaard (1984), Gough (1985) and Kurtz (1986).  Since all
three authors are well-known experts on both stellar oscillations and the
English language, I was inclined to accept Trimble's suggestion.  I have
since checked the original papers and found the `e' to be present in all
three, so the ADS made a mistake (which they have since corrected).  I am
happy to retract my suggestion, and refer the interested reader to Gough
(1996) for an eloquent and scholarly defence of the `e'.

Having tracked down the seminal paper on this subject, namely that by
Christensen-Dalsgaard (1984), I can do no better than quote his definition
of asteroseismology as ``the science of using stellar oscillations for the
study of the properties of stars, including their internal structure and
dynamics.''  Hipparcos luminosities are important because asteroseismology
is most powerful in constraining stellar models when the oscillation
frequencies are accompanied by more traditional observables.  I will
concentrate on three classes of stars for which Hipparcos parallaxes are
particularly relevant.

\section{$\delta$~Scuti variables}

The $\delta$~Scuti variables are A and F stars lying on or near the main
sequence, inside the classical instability strip.  Recent reviews include
those by Guzik (2000) and Handler (2000), as well as the proceedings of a
workshop dedicated to these stars (Breger \& Montgomery 2000).  Applying
asteroseismic techniques to multi-mode $\delta$~Scuti stars has been
severely impeded by our inability to identify which oscillation modes are
being observed.  This difficulty is due both to the mysterious tendency of
these stars to oscillate in only a small subset of possible modes, as well
as the substantial frequency shifts produced by their fast rotation.  The
availability of precise Hipparcos parallaxes for some of these stars has
not so far generated much progress in interpreting the complicated
oscillation spectra (although see Bedding, Kjeldsen, \&
Christensen-Dalsgaard 1998 for a discussion of $\kappa^{\scriptscriptstyle
2}$~Boo).

%One exception might be the multi-mode variable $\kappa^{\scriptscriptstyle
%2}$~Boo.  As pointed out by Bedding, Kjeldsen, \& Christensen-Dalsgaard
%(1998), the distance of 47.9\,pc derived by Frandsen et~al.\ (1995) from
%asteroseismology is in perfect agreement with the Hipparcos distance of
%$47.6 \pm 1.9$\,pc.  However, the frequency identifications by Frandsen
%et~al.\ did not allow for rotational splitting, despite the fact that this
%star is known to be a rapid rotator.  The accuracy of the asteroseismic
%parallax may therefore be coincidental.

High-amplitude $\delta$~Scuti stars (HADS) are a subclass for which
Hipparcos parallaxes have been extremely valuable.  These stars, which
constitute about 10\% of the total, pulsate in only one or a few modes
(McNamara 1997; Petersen 1998; Alcock et~al. 2000).  Hipparcos parallaxes
were used by H{\o}g \& Petersen (1997) to confirm HADS as otherwise normal
stars that follow standard stellar evolution, ruling out alternative
low-mass scenarios that had been proposed in the literature.  In
particular, they showed that the population~II field variable SX~Phe is
well described by standard stellar models and is very similar to the
variable blue stragglers in globular clusters.

Petersen \& H{\o}g (1998) used Hipparcos data to construct a
period-luminosity relation for HADS, which is in agreement with previous
relations derived for clusters.  HADS therefore show promise as distance
indicators to both the Galactic Bulge (Alcock et~al.\ 2000), as well as to
globular clusters and nearby dwarf galaxies (McNamara 2000).

\section{Rapidly oscillating Ap stars}

Lying in the same part of the H-R diagram as the $\delta$~Scuti variables,
these are chemically peculiar magnetic stars that oscillate in
high-overtone p-modes.  Matthews, Kurtz, \& Martinez (1999) have shown that
Hipparcos parallaxes for a sample of twelve roAp stars agree reasonably
well with parallaxes estimated from asteroseismology.  However, there are
systematic differences that imply that roAp stars are cooler than indicated
by H$\beta$ photometry, suggesting a need to lower the effective
temperature scale of Ap stars.  Hopefully, angular diameters from
interferometry will soon be able to resolve this issue.

\section{Solar-like oscillations}

The scientific importance of solar-like oscillations can be assessed from
the tremendous impact that helioseismology has had on our understanding of
the Sun.  It seems certain that observations of similar oscillations in
other stars will allow new and very strong tests of stellar evolution
models.  For example, stars only slightly more massive than the Sun are
thought to have a convective core for part of their main-sequence lives.
The associated mixing will clearly have a dramatic influence on the
evolution and lifetime of a star, and this will be visible in the
oscillation frequencies.

The search for analogues of the 5-minute solar oscillations in other stars
finally seems to be bearing fruit.  For discussion of a long series of
disappointing and tentative results, see reviews by Brown \& Gilliland
(1994), Kjeldsen \& Bedding (1995), Heasley et~al.\ (1996) and Bedding \&
Kjeldsen (1998).  Successes are now starting to appear, with velocity
variations in two stars showing excellent evidence for oscillations:
Procyon (Martic et~al.\ 1999; Barban et~al.\ 1999) and the G2 subgiant
$\beta$~Hyi (Bedding et~al.\ 2000).  The subsequent comparisons of observed
frequencies with model calculations will rely heavily on Hipparcos
parallaxes.  This has already been the case for the G0 subgiant $\eta$~Boo,
in which good evidence for oscillations was detected by Kjeldsen et al.\
(1995).  At the time, the observed frequencies agreed with the ground-based
parallax.  Since then, Hipparcos has given a factor of three improvement in
parallax precision and maintained the agreement (Bedding et al.\ 1998),
which is an important early success for asteroseismology of solar-like
stars.

The recent progress in ground-based observations, as well as results from
the star tracker on the otherwise-failed WIRE satellite (Buzasi et~al.\
2000), illustrate the potential of three upcoming space missions: MOST
(Matthews et~al.\ 2000), MONS (Kjeldsen, Bedding, \& Christensen-Dalsgaard
2000) and COROT (Baglin et~al.\ 1998).  The prospects for obtaining
oscillation measurements on a range of solar-type stars look very good,
which therefore makes it vital to obtain as much information about the
targets as possible.  Hipparcos parallaxes will play a key role, and I
strongly urge those planning the GAIA mission not to design out the
capability to observe the very brightest stars.

\acknowledgments For travel support, I thank the Australian Research
Council and the Science Foundation for Physics in the University of
Sydney.


\begin{references}

\reference Alcock, C., et al.\ 2000, ApJ, 536, 798

\reference Baglin, A., et~al.\ 1998, in IAU Symp.\ 185, New Eyes to See
Inside the Sun and Stars, ed.\ F.-L.\ Deubner, J. Christensen-Dalsgaard \&
D.~W. Kurtz (Dordrecht: Kluwer), 301 ({\tt
http://www.astrsp-mrs.fr/projets/corot/})

\reference Barban, C., {Michel}, E., {Martic}, M., {Schmitt}, J., {Lebrun},
J.~C., {Baglin}, A., \& {Bertaux}, J.~L. 1999, A\&A, 350, 617

\reference Bedding, T.~R., Butler, R.~P., Kjeldsen, H., Baldry, I.~K.,
O'Toole, S.~J., Tinney, C.~G., Marcy, G.~W., Kienzle, F., \& Carrier, F.
2000, submitted to ApJ Letters

\reference Bedding, T.~R., \& Kjeldsen, H. 1998, in
 ASP Conf.\ Ser.\ Vol.\ 154, 
 Tenth Cambridge Workshop on Cool Stars, Stellar Systems and the Sun,
 ed R.~A. Donahue \& J.~A. Bookbinder
 (San Francisco: ASP),
 301

\reference Bedding, T.~R., Kjeldsen, H., \& Christensen-Dalsgaard, J.,
1998, in 
 ASP Conf.\ Ser.\ Vol.\ 154, 
 Tenth Cambridge Workshop on Cool Stars, Stellar Systems and the Sun,
 ed R.~A. Donahue \& J.~A. Bookbinder
 (San Francisco: ASP),
 CD--741
 ({\tt astro-ph/9709005})

\reference
Breger, M., \& Montgomery, M. 2000,
 ASP Conf.\ Ser.\ Vol.\ 210,
 Sixth Vienna Workshop in Astrophysics: Delta Scuti and Related Stars

\reference Brown, T.~M., \& Gilliland, R.~L. 1994, ARA\&A, 33, 37

\reference Buzasi, D.~L., Catanzarite, J., Laher, R., Conrow, T., Shupe,
D., Gautier~III, T.~N., \& Kreidl, T. 2000, ApJ, 532, L133

\reference Christensen-Dalsgaard, J. 1984, in 
 Workshop on Space Research in Stellar Activity and Variability,
 ed A. Mangeney \& F. Praderie
 (Meudon: Observatoire de Paris),
 11

\reference Frandsen, S., Jones, A., Kjeldsen, H., Viskum, M., Hjorth, J.,
Andersen, N.~H., \& Thomsen, B. 1995, A\&A, 301, 123

\reference Gough, D.~O. 1985, Nature, 314, 14

\reference Gough, D.~O. 1996, The Observatory, 116, 313.  Erratum: 117, 72

\reference Guzik, J.~A. 2000, in 
Variable Stars as Essential Astrophysical Tools,
ed C. Ibanoglu
(Dordrecht: Kluwer),
213 

\reference Handler, G. 2000, in
 ASP Conf.\ Ser.\ Vol.\ 203,
 IAU Coll.\ 176: The Impact of Large-Scale Surveys on Pulsating Star
   Research,
 ed L. Szabados \& D. Kurtz
 (San Francisco: ASP),
 408

\reference Heasley, J.~N., Janes, K., Labonte, B., Guenther, D., Mickey,
D., \& Demarque, P. 1996, PASP, 108, 385

\reference H{\o}g, E., \& Petersen, J.~O. 1997, A\&A, 323, 827

\reference Kjeldsen, H., \& Bedding, T.~R. 1995, A\&A, 293, 87

\reference Kjeldsen, H., Bedding, T.~R., Viskum, M., \& Frandsen, S. 1995,
AJ 109, 1313

\reference Kjeldsen, H., Bedding, T.~R., \& Christensen-Dalsgaard, J.,
2000, in
 ASP Conf.\ Ser.\ Vol.\ 203,
 IAU Coll.\ 176: The Impact of Large-Scale Surveys on Pulsating Star
   Research,
 ed L. Szabados \& D. Kurtz
 (San Francisco: ASP),
 73 ({\tt http://astro.ifa.au.dk/MONS})

\reference Kurtz, D.~W. 1986, in 
 IAU Symp.\ 118,
 Instrumentation and Research Programmes for Small Telescopes,
 ed J.~B. Hearnshaw \& P.~L. Cottrell
 (Dordrecht: Reidel),
 251

\reference Martic, M., Schmitt, J., Lebrun, J.-C., Barban, C., Connes, P.,
Bouchy, F., Michel, E., Baglin, A., Appourchaux, T., \& Bertaux, J.-L.,
1999, A\&A, 351, 993

\reference Matthews, J.~M., {Kurtz}, D.~W., \& {Martinez}, P. 1999, \apj,
511, 422

\reference Matthews, J.~M. et al. 2000, in
 ASP Conf.\ Ser.\ Vol.\ 203,
 IAU Coll.\ 176: The Impact of Large-Scale Surveys on Pulsating Star
   Research,
 ed L. Szabados \& D. Kurtz
 (San Francisco: ASP),
 74 ({\tt http://www.astro.ubc.ca/MOST})

\reference McNamara, D. 1997, PASP, 109, 1221

\reference McNamara, D.~H. 2000, PASP, 112, 1096

\reference Petersen, J.~O. 1998, in 
The First MONS Workshop: Science with a Small Space Telescope,
ed H. Kjeldsen \& T.~R. Bedding
(Aarhus: Aarhus Universitet),
119

\reference Petersen, J.~O., \& H{\o}g, E. 1998, A\&A, 331, 989

\reference Trimble, V. 1995, PASP, 107, 1012


\end{references}
\end{document}